\documentclass[aps,prl,superscriptaddress,twocolumn,showpacs,nofootinbib]{revtex4-2}
\usepackage{changes} 

\usepackage{graphicx}
\usepackage{subfigure}
\usepackage{amsthm}
\usepackage{amsmath}
\usepackage{verbatim}
\usepackage{dcolumn}
\usepackage{bm}
\usepackage{float}
\usepackage{color}
\usepackage[colorlinks=true,citecolor=blue,linkcolor=blue,urlcolor=blue]{hyperref}%
\usepackage{xcolor}
\usepackage{dsfont}
\usepackage{longtable}
\usepackage{tabularx}
\usepackage[ruled,vlined]{algorithm2e}
\allowdisplaybreaks
\newcommand{\floor}[1]{\lfloor {#1} \rfloor}
\usepackage{microtype}

\begin{document}
\title{Toward Quantum analog Simulation of Many-Body Supersymmetry with Rydberg Atom Arrays}
\date{\today}

\author{Hrushikesh Sable}
\email{hsable@vt.edu}
\affiliation {Department of Physics, Virginia Tech, Blacksburg, Virginia 24061, USA}
\author{Nathan M. Myers}
\affiliation{Department of Physics, Virginia Tech, Blacksburg, Virginia 24061, USA}
\author{Vito W. Scarola}
\affiliation {Department of Physics, Virginia Tech, Blacksburg, Virginia 24061, USA}

\begin{abstract}
A topological quantum number, the Witten index, characterizes supersymmetric models by probing for zero energy modes and the possibility of supersymmetry breaking. We propose an averaging method to infer the Witten index in quantum analog simulators. Motivated by recent work on Rydberg atoms trapped in optical tweezer arrays, we consider a related supersymmetric XXZ spin model. We show how to infer the Witten index from open-system averaging and numerically demonstrate its topological robustness in this model. Our Letter defines a route for quantum analog simulators to directly identify many-body topological physics. 
\end{abstract}

\maketitle

\emph{Introduction$-$}Theories with supersymmetry (SUSY) were first constructed to generalize relativistic quantum field theories to help address issues within the standard model of particle physics \cite{Aitchison_2007,witten1981}.
SUSY methods were then applied to make important contributions to the study of nonrelativistic quantum systems, e.g., SUSY breaking mechanisms \cite{Nicolai_1976, witten_1982_constraints,cooper1995,junker_1996_supersymmetric}. In particular, $\mathcal{N}=2$ SUSY Hamiltonians can be written as $\hat{H}=\hat{Q}^\dagger \hat{Q}+\hat{Q} \hat{Q}^\dagger$,  where $\hat{Q}$ and $\hat{Q}^{\dagger}$ are nilpotent, non-Hermitian supercharges.  $\hat{H}$ comes with an operator $\hat{F}$ defined by the commutators: $[\hat{F},\hat{Q}]=-\hat{Q}$ and $[\hat{F},\hat{Q}^{\dagger}]=\hat{Q}^{\dagger}$.  Such models have important properties \cite{witten_1982_constraints}:  all energy eigenvalues satisfy $E\geq0$ and all $E>0$ states have pairwise degenerate SUSY partners sorted by the even or odd parity of $\hat{F}$ eigenvalues. SUSY is unbroken when $\hat{H}$ has at least one $E=0$ state, otherwise it is spontaneously broken. For supercharges constructed from nonlocal symmetries, the existence of zero-energy states has crucial implications for underlying topological physics.

For $\mathcal{N}=2$ SUSY, there are only two topological indices based on $\hat{F}$: Tr[$(-1)^{\hat{F}}]$ \cite{witten_1982_constraints} and Tr[$\hat{F}(-1)^{\hat{F}}]$ \cite{CECOTTI1992}. We focus on the regularized Witten index \cite{witten_1982_constraints}:
\begin{equation}
    W = {\rm Tr }\left[ (-1)^{\hat{F}} {\rm exp}(-\beta_0 \hat{H}) \right]
    \label{math_witten_ind}
\end{equation}
 which was designed as a mathematical probe of spontaneous breaking of SUSY, where $\beta_0 \geq 0$ is a cutoff parameter. 
$W$ is a spectral topological \cite{Bernstein1985} index (equivalent to the Euler characteristic \cite{junker_1996_supersymmetric}).  $W$ identifies $E=0$ states since the trace over $E>0$ states vanishes for any $\beta_0$, leaving $W$ to be nonzero only if there are $E=0$ state(s). 
Furthermore, finite-sized system trends in $W$ persist to the thermodynamic limit, thus demonstrating its utility for finite-size simulation \cite{witten_1982_constraints} compatible with existing technology. 
In addition, $W$ is well defined in strongly correlated SUSY models, making it an intrinsically useful probe of quantum many-body models which are otherwise nontrivial to characterize.

The Witten index can be indirectly related to known models of quantum matter.  For instance, $W$ is related to observables in noninteracting models of two-dimensional Pauli paramagnets \cite{junker_1996_supersymmetric,AHARONOV1979} (where $\hat{F}$ is magnetization) and translationally invariant Majorana fermions (where $\hat{F}$ is particle number) \cite{Hsieh2016}.  Exact SUSY is also known to exist in certain strongly correlated XXZ spin chains \cite{fendley_2003_lattice, yang_2004_nonlocal} (where $\hat{F}$ is a combination of magnetization and chain length), as discussed below. Additionally, SUSY is studied in Rabi models in quantum optics \cite{tomka_2015_supersymmetry}, where $W$ is related to the parity of excitations, and in Bose-Fermi mixtures \cite{yu2008}. 

Computationally, determining $W$ lies in the hardest complexity class, number P complete \cite{crichigno2020}. With limited analytical techniques for quantum many-body systems, analog simulation would provide a promising alternative route to study $W$. However, despite its broad applicability, both in high-energy and condensed matter systems, $W$ remains a mathematical tool unrelated to an observable obtained from a physical averaging process and lacks a general protocol connecting it to experiment.

The disconnect between $W$ and experiments has become more pressing with the advent of atomic, molecular, and optical (AMO) quantum analog simulators \cite{Lewenstein2007,Bloch2008,Saffman2010,browaeys_2020_manybody,WU2021a,Monroe2021}. Recently, ion trap experiments have implemented a non-interacting SUSY model \cite{CAI2022} and it has been demonstrated that Rydberg atoms trapped in optical lattices can probe kink dynamics of a SUSY model of interacting fermions \cite{MINAR2022a}, as well as the emergent SUSY at critical points \cite{dalmonte2015, mikheil2022, fromholz2022}. SUSY-based cooling with neutral atoms in optical tweezers has also been proposed \cite{Luo2021}. In particular, atoms in optical tweezer arrays show immense potential for quantum simulation, with recent experiments demonstrating arrays with up to 6100 atoms \cite{MANETSCH2024} and hold times on the order of hours \cite{SCHYMIK2021,GYGER2024}.  Furthermore, atomic levels defining pseudospins allow these arrays to effectively simulate a large class of many-body spin models \cite{browaeys_2016_experimental, whitlock2017,NGUYEN2018,GEIER2021a,Ebadi2021,Scholl2021,signoles2021,BLUVSTEIN2022,scholl2022,BLUVSTEIN2024}.  

Here, we propose a general method for quantum analog simulation of many-body SUSY physics. We address the issue of experimentally detecting the signatures of $W$ by defining a normalized Witten index, $\widetilde{W}$, using an average with an engineered energy density, $\hat{\sigma}$. If $\hat{\sigma}$ samples SUSY partners in an unbiased manner, it reveals key aspects of $W$ while directly connecting to physical averaging. We apply our proposed method to many-body SUSY in an XXZ spin model \cite{yang_2004_nonlocal} as a test bed.  We show that Rydberg atoms in tweezer arrays are surprisingly relevant for quantum analog simulation of this SUSY model with ongoing experiments \cite{browaeys_2016_experimental, whitlock2017, NGUYEN2018,GEIER2021a,Ebadi2021,Scholl2021,signoles2021,BLUVSTEIN2022,scholl2022,BLUVSTEIN2024}. 

Probing nonlocal quantities at (or near) integrable points, e.g., $W$ in SUSY models, makes achieving closed-system thermalization challenging \cite{DALESSIO2016}.  We therefore adopt an open-system framework motivated by quantum collisional models \cite{Barra2017, Ciccarello2022, rodrigues2019,TabaneraBravo2023,myers2024} to test the observability of $\widetilde{W}$ using numerical simulations. This open-system averaging protocol should, for any SUSY theory, be grand canonical with respect to fluctuations in $\hat{F}$, ensuring sampling of both SUSY partners for an unbiased evaluation of $\widetilde{W}$.  
As can be seen from the example SUSY models discussed, in general, $\hat{F}$ depends on particle number and/or system size, and calculating $\widetilde{W}$ involves grand canonical averaging (GCA) over varying system sizes or particle number, which is nontrivial for analog simulators as these are nearly isolated systems. Consequently, we introduce an experimentally suitable method of approximating GCA by sampling over fixed-size canonical averages. 
This approximation works at low temperatures as the size-changing system-reservoir interactions can be neglected. 
This approach is helpful for GCA with closed AMO simulators such as Rydberg atoms \cite{browaeys_2020_manybody}, neutral atoms in optical lattices \cite{gross2017} and ion traps \cite{Monroe2021}.
We numerically demonstrate that $\widetilde{W}$ displays remarkable topological robustness. Our method sets the stage for quantum analog simulators to quantitatively study many-body topological effects in SUSY models and study conditions under which SUSY breaks. 

\begin{figure}[t]
   \includegraphics[width=\linewidth]{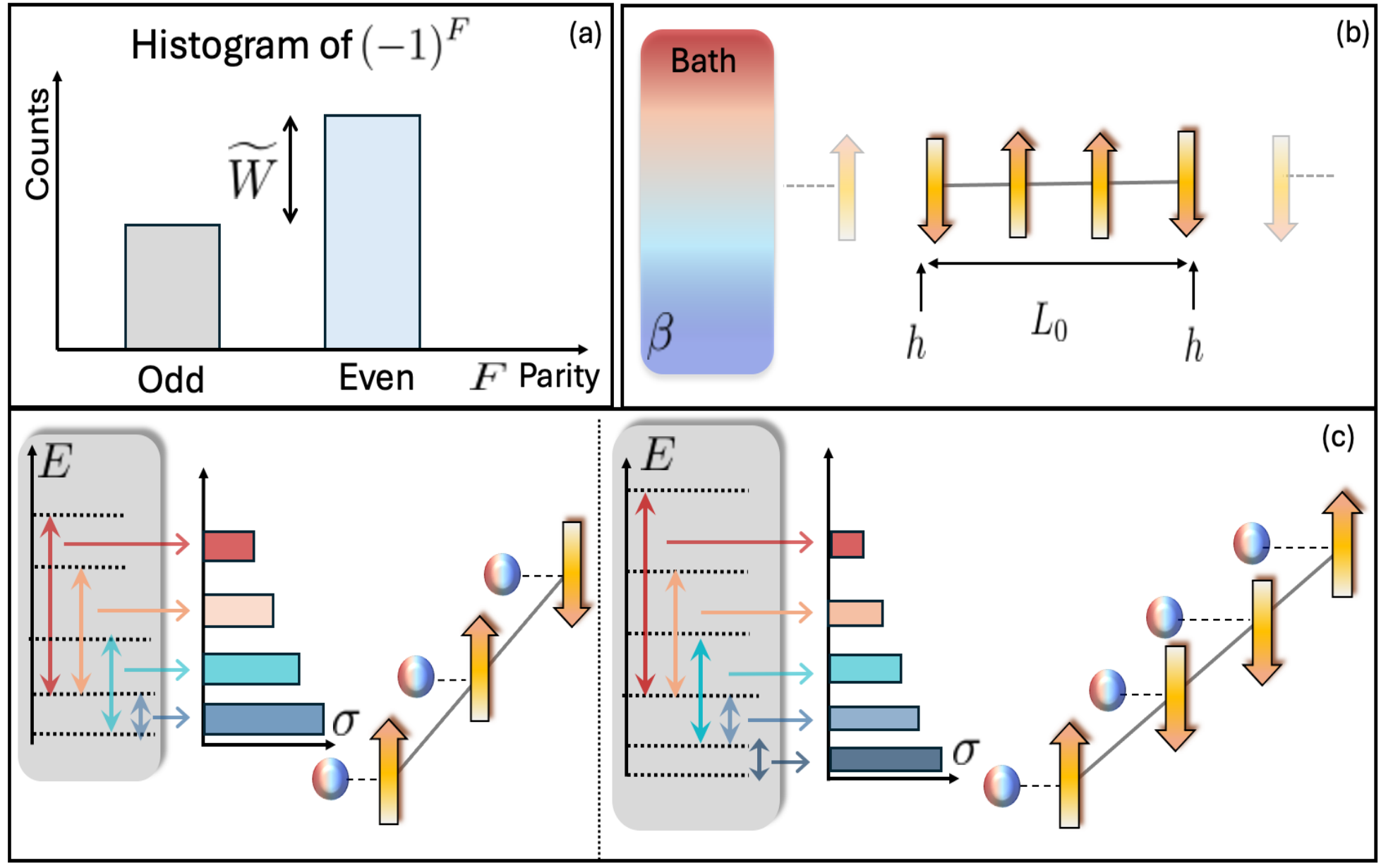}
    \caption{ 
    (a) A schematic histogram of measurements of the parity of $\hat{F}$ used to build $\widetilde{W}$, generic to any SUSY model. (b) GCA: The system consists of a generic spin chain of variable length coupled to a bath at inverse temperature $\beta$, with color gradient from blue to red for increasing energies. The interaction with the bath changes the length \emph{and} energy of the chain. One such spin chain, of size $L_0$, is shown. (c) Quasigrand canonical averaging (QGCA):  Spin chains of different sizes, each interacting individually with ancillae drawn from a thermal ensemble, schematically shown by a histogram representing the probability distribution over the chain's excitation energies depicted by arrows of colors consistent with (b). At each interaction, an ancilla corresponds to one of these energies. }
    \label{schm}
\end{figure}

\emph{Normalized Witten index$-$} We propose a counterpart to Eq.~\eqref{math_witten_ind}, applicable to quantum simulation of any SUSY model $\hat{H}$.  If $\hat{F}$ is an observable, we define$-$
\begin{equation}
    \widetilde{W} = {\rm Tr }\left[ (-1)^{\hat{F}} \hat{\sigma} \right],    
    \label{obser_witten_ind}
\end{equation}
where $\hat{\sigma}$ is a normalized distribution with weights $\sigma$, such that (i) $\hat{\sigma}$ yields weight at $E=0$ and (ii) $\hat{\sigma}$ has equal sampling probability of some $E>0$ SUSY partners. A general distribution over system eigenstates can bias one degenerate partner over another \cite{supplemental}, necessitating (ii).
An energy distribution meeting these conditions can be engineered by the system exchanging energy with a reservoir through repeated interactions. As a convenient example, we consider dynamics that drive $\hat{\sigma}$ to the thermal distribution, ${\rm exp}(-\beta \hat{H})/\mathcal{Z}$, where $\beta$ is the inverse temperature of a reservoir and $\mathcal{Z}$ is the partition function. However, thermalization is not a requirement for $\hat{\sigma}$, and any steady state distribution meeting the above conditions would work.

Once $\hat{F}$ is identified and its eigenvalues observed, averaging its parity $(-1)^{\hat{F}}$ using $\hat{\sigma}$ yields $\widetilde{W}$.  Counting this parity naturally excludes excited states in SUSY models since they cancel in $\widetilde{W}$.  But the presence of an asymmetry in counts will indicate the presence of one or more zero-energy modes and reflects in $\widetilde{W}$, as shown in Fig.~\ref{schm}(a). Figures~\ref{schm}(b) and \ref{schm}(c) show schematics of two different averaging protocols for engineering $\hat{\sigma}$, as described below.

\emph{SUSY XXZ model$-$} To demonstrate our method we consider the XXZ Hamiltonian of an $L$-site spin chain \cite{yang_2004_nonlocal}: 
\begin{eqnarray}
    \hat{H}_{\rm XXZ} &=& \sum_{i=1}^{L-1} \left[J(\hat{S}_i^+ \hat{S}_{i+1}^- + {\rm H.c. } ) + \Delta \hat{S}_i^{z} \hat{S}_{i+1}^z\right] \nonumber \\ & -& h \left(\hat{S}_1^z + \hat{S}_L^z \right) + \left(3L - 1 \right)/4,
    \label{xxz_susy}
\end{eqnarray}
where $\hat{S}_{i}^{+} (\hat{S}_{i}^{-})$ are the spin $1/2$ raising (lowering) operators and $\hat{S}_i^{z}$ is the $z$ component of the spin at site $i$.  We consider open boundaries such that edge sites experience the energy shift $h$.  The last term is an energy cost to add sites and serves as a chemical potential. 
The Hamiltonian in Eq.~\eqref{xxz_susy} exhibits SUSY at $(J,\Delta, h) = (-1,1,1/2)$ \cite{fendley_2003_lattice, yang_2004_nonlocal}.

\begin{figure}[h]
   \centering
    \includegraphics[width=0.8\linewidth]{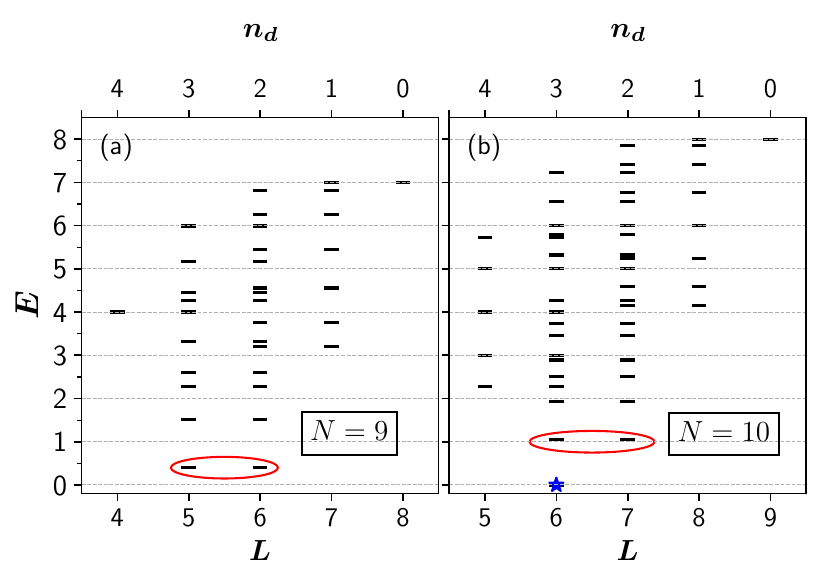}
   \caption{Energy of Eq.~\eqref{xxz_susy} versus length at the SUSY point for fixed $N$.  Panel (a) shows the $N=9$ case with no $E=0$ state, and thus $W=0$.
   Panel (b) shows the case with the $E=0$ mode present (star) for $N=10$ and therefore $W \neq 0$.  
   The lowest SUSY partners with opposite parities determined by $n_d$ are encircled.}
   \label{energy_spec}
\end{figure}

Equation~\eqref{xxz_susy} is an excellent approximation to recent experiments realizing the XXZ model using Rydberg atoms with resonant dipole-dipole interactions trapped in optical tweezer arrays \cite{browaeys_2016_experimental, whitlock2017, signoles2021,GEIER2021a,scholl2022}.  Typical spin-spin interaction
energy scales are $\sim 1-100 \hspace{0.1cm}\hbar\times\text{kHz}$.  A differential van der Waals shift between interacting Rydberg atoms leads to an intrinsic edge bias which can be modeled as an edge magnetic field $h$ \cite{NGUYEN2018,signoles2021}, allowing for the physical realization of Eq.~\eqref{xxz_susy} with existing Rydberg technologies \cite{browaeys_2016_experimental, whitlock2017,barredoThreeDimensionalTrappingIndividual2020, browaeys_2020_manybody, Scholl2021, signoles2021,Ebadi2021,GEIER2021a,scholl2022}. 

Figure~\ref{energy_spec} illustrates the SUSY spectra of $H_{\rm XXZ} $.  For this model,
\begin{align}
\hat{F} = \bigg (L-\sum_{i=1}^L \hat{S}_i^z \bigg)/2.
\end{align}
The eigenvalues of $\hat{F}$ are the number of down spins, $n_d$.
Degenerate SUSY eigenstates correspond to chains of different $L$ and have an even or odd $n_d$.  The nonlocal supercharges for this model take a state with quantum numbers $L$ and $n_d$ to one with $L+1$ and $n_{d}-1$ thus motivating the conserved quantum number \cite{yang_2004_nonlocal}:
  \begin{align}
N \equiv L + n_d + 1.
\end{align}
Using $N$, one finds two types of sectors: $N=3j$ and $N\neq 3j$, for any integer $j$.  
In general, the spectrum has only one $E=0$ state for every $N\ne 3j$, and no $E=0$ state when $N=3j$.  This model therefore has unbroken SUSY and $W=(-1)^{\floor{N/3}}$ 
\cite{ yang_2004_nonlocal}.  Consequently, determining $W$ for finite sized systems provides information about the thermodynamic limit \cite{witten_1982_constraints}. This property makes $\widetilde{W}$ particularly accessible with existing quantum simulation technology as very large system sizes are not required.

\emph{Open system dynamics$-$} We consider an open system framework that generates the target distribution $\hat{\sigma}$ through a discretized time evolution. The physical intuition for this approach comes from collisional models where open system dynamics is modeled as a series of discrete interactions between the system and ancillae forming a reservoir \cite{Ciccarello2022, Barra2017, rodrigues2019, arisoy2019, Strasberg2017, TabaneraBravo2023}. 

Recent work on ancilla atoms for midcircuit measurements on Rydberg-based qubits \cite{BETEROV2015,GLAETZLE2017,DEIST2022,SINGH2022,GRAHAM2023,LIS2023,MA2023,NORCIA2023,ANAND2024} can be leveraged to create reservoirs for many-body quantum analog simulation \cite{Schonleber2018, METCALF2020}.
These setups enable independent driving of ancilla qubits that can be adapted to simulate open system dynamics.
In our Letter, the ancilla-system interactions are approximated using a Monte Carlo Metropolis method \cite{metropolis1953, myers2024},
where every time step corresponds to creating or annihilating an excitation of energy resonant with the energy gaps of Eq.~\eqref{xxz_susy}. This approximation holds provided the scattering time between the reservoir and system, $\tau$, occurs much faster than system dynamics, i.e., $\tau \ll \vert \hbar/J \vert$. We employ two averaging protocols, one that is ideal and one that is approximate but experimentally relevant. 
 
Our first protocol is GCA, shown in Fig.~\ref{schm}(b), where the system-reservoir interactions change both $E$ \emph{and} $L.$
While excitations that vary the size of the chain are difficult to realize experimentally, this method serves to provide an ideal reference case. It yields the distribution $\hat{\sigma}$ with weights $\sigma(E)$, dependent only on $E$, thereby satisfying condition (ii). 
 
\begin{figure}[h]
    \includegraphics[width=\linewidth]{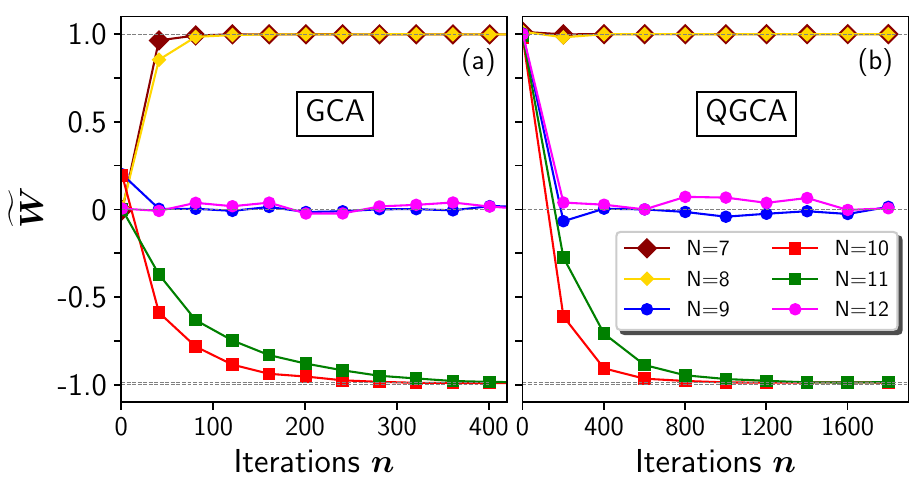}
    \caption{$\widetilde{W}$ as a function of iterations $n$ for different $N$ sectors at $\beta=5$. Each iteration can be interpreted as a time step of duration $O(1/J)$ in the XXZ model.
     The steady-state values agree with the equilibrium Gibbs distribution values depicted by dashed lines. Note that $N=9$ (blue) and $N=12$ (magenta) cases average to zero. The largest absolute error in the steady-state values is of the order $O(10^{-3})$ in (a) and $O(10^{-2})$ in (b). The results are generated by averaging over $50000$ Monte Carlo runs \cite{supplemental}.} 
    \label{wittn_iter_b5}
\end{figure}

Our second approach is QGCA, which approximates the length-changing system-reservoir interactions but is more experimentally accessible.
We consider a set of spin chains of different $L$, each interacting with a dissipative environment composed of ancillae, as shown in Fig.~\ref{schm}(c). The ancillae energies are engineered such that the system-ancilla interactions induce transitions among the many-body energy eigenstates of the spin chain, driving each chain towards its canonical thermal state. We average over the canonical distributions for different length chains in the ensemble. This approach is \textit{quasigrand} canonical, as variations in $L$ are not generated by the interaction with the bath.   

One recently proposed scheme to produce canonical thermalization dynamics uses damped, driven ancillae weakly coupled to the spin system \cite{METCALF2020}. The ancillae energy spacing is swept across the system's many-body spectrum, avoiding fine-tuning the ancilla spacing to each gap. At most, $L$ ancillae are required for thermalization, but less may be required if the system-ancilla couplings produce ergodic dynamics \cite{METCALF2020}. 

Estimating $\widetilde{W}$ with QGCA is a good approximation to GCA, provided that the reservoir temperature is below the system energy gap to the ground state. Since the canonical partition functions for all the system sizes approach unity at low temperatures \cite{supplemental}, the degenerate eigenstates are sampled with equal probability.
Note that sampling over the eigenstates of individual chains itself leads to the length-dependent weights $\sigma(E,L)$, violating (ii). However, at low temperatures, $\widetilde{W}$ computed using $\sigma(E,L)$, when summed over different $L$, is in excellent agreement with the estimate of $\widetilde{W}$ computed using $\sigma(E)$. This approximation breaks down as the temperature approaches the energy gap \cite{supplemental} resulting in a distribution that does not equally sample SUSY partners.
This method of approximating the GCA with the QGCA is general, provided temperature is low enough that the spectrum below it is nearly identical across system sizes \cite{supplemental}.
  
\emph{Dynamics and estimates of $\widetilde{W}-$} We numerically compare steady state estimates of $\widetilde{W}$ using both GCA and QGCA.  Figure ~\ref{wittn_iter_b5}(a) plots the dynamics of $\widetilde{W}$ for different $N$ sectors using GCA.  For $N=3j$, the energy spectrum has no $E=0$ eigenstates, and hence $W = \widetilde{W} = 0$ [Fig.~\ref{energy_spec}(a)].  Our simulations yield this expected result in the steady-state regime. For $N \ne 3j$, we also find that $\widetilde{W}$  reduces to $W = (-1)^{\floor{N/3}}$ at long times (dashed lines). See Ref.~\onlinecite{supplemental} for lower $N$ values.

Figure~\ref{wittn_iter_b5}(b) shows the same as (a) but with QGCA and we see both results converge to the same value. The $E=0$ mode in the spectrum of every chain in the ensemble makes the partition functions of different system sizes equal at low temperatures. However, if the temperature is increased above the energy gap, considerable differences between the two approaches arise \cite{supplemental}. To maintain the accuracy of the QGCA, we require that the temperature decreases as system size increases \cite{supplemental}. However, since $\widetilde{W}$ for this model does not scale with system size, a small-scale simulation not requiring such low temperatures is sufficient to determine $\widetilde{W}$, once the $(-1)^{\floor{N/3}}$ pattern emerges.

We find that, at temperatures below the gap, the Witten index can be inferred from QGCA. 
The QGCA technique physically corresponds to the steady-state measurements of $n_d$ on separate spin chains (of different $L$), and tabulating the outcomes of the parity as shown in Fig.~\ref{schm}(a). These measurement results can then be combined to find $\widetilde{W}$.
Note that the $\widetilde{W}$ extracted from a distribution at high temperature, respecting both (i) and (ii), is equivalent to the one extracted under QGCA at low temperature, up to the normalization $\mathcal{Z}$.

Implementing our protocol in an experiment requires careful consideration of the timescales. In Rydberg platforms, the XXZ model is engineered through external driving. For instance, fast driving fields are needed to establish a rotating frame in Ref.~\onlinecite{NGUYEN2018} or Floquet pulses in  Ref.~\onlinecite{scholl2022}. 
We assume that these driving fields set the shortest time scale $\lesssim J^{-1}$. These fast-driving fields must be interleaved with system-reservoir interactions that, as we find, take a suitably long time to implement thermalization, $\sim 10^2 J^{-1}$. Finally, we assume that external sources of decoherence of the many-body quantum state occur on the longest timescales.

\begin{figure}[t]
    \centering
    \includegraphics[width=\linewidth]{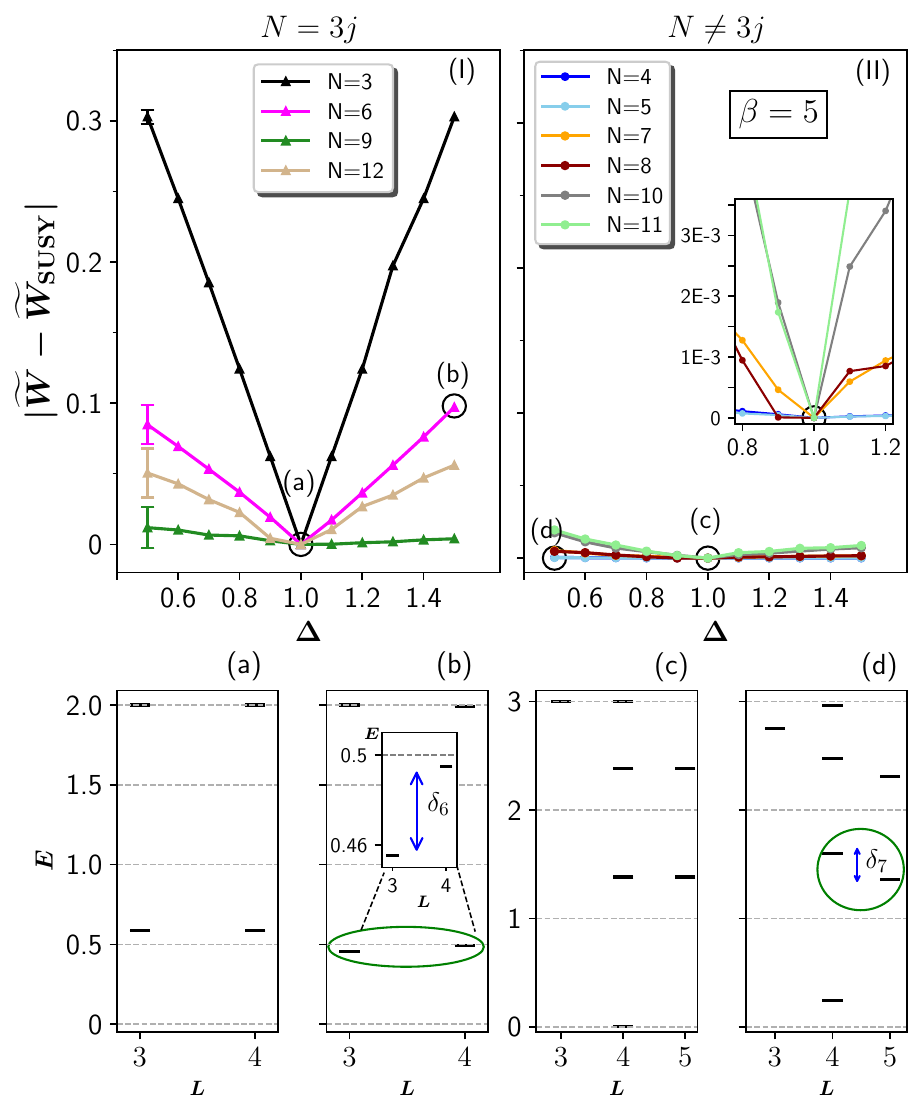}  
    \caption{$|\widetilde{W} - \widetilde{W}_{\rm SUSY}|$ as $\Delta$ is varied across the SUSY point, for $\beta = 5$, for two cases: $N=3j$ (I) and $N \ne 3j$ (II). (II) has an $E=0$ state and demonstrates topological protection.  Characteristic error bars in the $N=3j$ case are shown, while for $N \ne 3j$ case, they are comparable to the point size.
    The inset in (II) shows the variation close to the SUSY point obtained with $50000$ samples. Panels (a) and (b) illustrate a part of the energy spectrum at $\Delta = 1$ and $1.5$ respectively, for $N=6$, while (c) and (d) show the spectrum at $\Delta=1.0$ and $0.5$ respectively, for $N=7$. The zoomed inset in (b) shows the splitting $\delta_6 \approx 3.9\times 10^{-2}$. } 
    \label{nonsusy_b5}
\end{figure}

\emph{Topological protection-} We now demonstrate topological protection of $\widetilde{W}$.  We use GCA for the results in this section (See Ref.~\onlinecite{supplemental} for results using QGCA). Such protection implies that quantum simulators can study SUSY in the XXZ model even in the presence of imperfections in the Hamiltonian parameters about the SUSY point. 

Figure~\ref{nonsusy_b5} depicts $|\widetilde{W}-\widetilde{W}_{\rm SUSY}|$ as $\Delta$ is varied across the SUSY point. For the $N=3j$ case, away from the SUSY point, $\widetilde{W}$ is nonzero due to degeneracy breaking. In Fig.~\ref{nonsusy_b5} (I) we find a nonvanishing slope in the linear dependence of $|\widetilde{W}-\widetilde{W}_{\rm SUSY}|$ on $\Delta$. To see this, we expand $|\widetilde{W}-\widetilde{W}_{\rm SUSY}|$ up to first order in $\Delta - 1$:
\begin{equation}
|\widetilde{W}-\widetilde{W}_{\rm SUSY}|  = c_N\beta |\Delta-1|,
 \label{brokeSUSY}
 \end{equation}
where $c_N$ is a numerical factor proportional to the slope of the degeneracy splitting $\delta_N$ with $\Delta$.  To see the splitting we compare part of the spectrum at (away from) the SUSY point in  Fig.~\ref{nonsusy_b5}(a) [Fig.~\ref{nonsusy_b5}(b)]. See Ref.~\onlinecite{supplemental} for discussion over finite size effects and degeneracy splitting.

Figure~\ref{nonsusy_b5} (II) plots the case $N \neq 3j$ where the zero energy mode leads to topological protection of $\widetilde{W}$. We find $|\widetilde{W}-\widetilde{W}_{\rm SUSY}|$ to be suppressed by orders of magnitude compared to the $N=3j$ case, implying robustness. To illustrate this case, the energy spectrum 
at [Fig.~\ref{nonsusy_b5}(c)] and away [Fig.~\ref{nonsusy_b5}(d)] from the SUSY point is shown. 

Figure~\ref{nonsusy_b5}(d)  shows the zero energy mode shifted to a nonzero value.  However, at low temperatures, this nondegenerate mode has nearly a unity probability, and 
$\widetilde{W}$ remains insensitive to the breaking of the degeneracy of higher energy modes.  The first order expansion is$-$
\begin{equation}
 |\widetilde{W}-\widetilde{W}_{\rm SUSY}| = c_N \beta {\rm exp}(-\beta E_1) |\Delta-1|,
 \label{unbrokeSUSY}
\end{equation}
where $E_1$ is the lowest $E>0$ state in the SUSY spectrum \cite{supplemental}.  
Comparing Eqs.~\eqref{brokeSUSY} and ~\eqref{unbrokeSUSY}, we see that the factor ${\rm exp}(-\beta E_1)$ is responsible for the vanishing slope and thus the topological protection, thereby explaining the negligible variation in $\widetilde{W}$ at low temperatures.  
 
Our numerical results of $|\widetilde{W}-\widetilde{W}_{\rm SUSY}|$ are in good agreement with the analytical results in Eqs.~\eqref{brokeSUSY} and \eqref{unbrokeSUSY} \cite{supplemental}.  A similar topological protection of $\widetilde{W}$ manifests when $J$ is varied \cite{supplemental}.  But we find that topological protection breaks down for temperatures near the gap, as is expected from Eq.~\eqref{unbrokeSUSY} \cite{supplemental}. This topological protection demonstrates that the SUSY is robust against fluctuations provided they are smaller than the gap.

\emph{Summary and outlook$-$} Symmetries encoded in supercharges can connect to observables, $\hat{F}$, which, when processed with appropriate averaging protocols, reveals SUSY.  We proposed a normalized Witten index as an observable and have constructed a corresponding averaging protocol.  We studied the protocol in an XXZ spin model relevant for ongoing experiments with Rydberg atom arrays.  Through our open-system numerical simulations we have demonstrated observability of the normalized Witten index and topological protection arising from zero energy modes. 

\begin{acknowledgments}
We acknowledge support from AFOSR-FA9550-23-1-0034,FA9550-19-1-0272 and ARO-W911NF2210247. 
\end{acknowledgments}

\bibliography{references}
		
\end{document}


\title{Supplementary material for ``Toward Quantum Analog Simulation of Many-Body Supersymmetry with Rydberg Atom Arrays"}

\maketitle

\section{Conditions on distribution $\hat{\sigma}$}
\begin{figure}[h]
   \centering
    \includegraphics[width=0.5\linewidth]{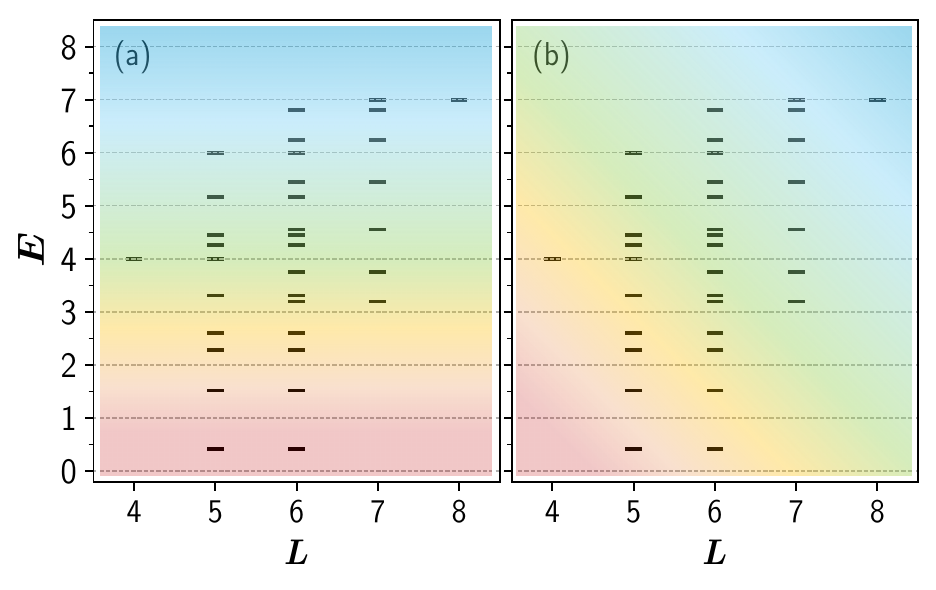} 
   \caption{Energy spectrum of the SUSY XXZ model at the SUSY point for $N=9$. The background color schematically represents a distribution over the eigenstates. Panel (a)\big((b)\big) illustrates the case when the distribution respects (violates) condition $(ii)$ listed in the main text.
   }
   \label{spec_grad}
\end{figure}

In the main text, we have discussed the two conditions that $\hat{\sigma}$ needs to satisfy to be used in the computation of $\widetilde{W}$. In this section, we discuss the reason why these conditions are required. 
If the distribution depends only on the eigenstates of the SUSY Hamiltonian, then by virtue of SUSY, the weights of the degenerate partners are equal as $\sigma = \sigma(E)$, which automatically satisfies condition $(ii)$. This is shown in Fig.~\ref{spec_grad}a, where the color gradient in the background of the energy spectrum schematically represents the weights of the distribution. However, in general, a distribution over the system eigenstates can bias a certain degenerate partner over another. This happens when the weights in the distribution depend on parameter $\alpha$ in addition to eigenvalues, $\sigma = \sigma(E, \alpha)$. This is illustrated in Fig.~\ref{spec_grad}b, where the $\alpha = L$. That is, the two degenerate eigenstates do not have same weights, owing to different $L$. The possibility of such general distributions necessitates condition $(ii)$. 

\section{Computational method}\label{methodology}
This section focuses on the computational methodology used in the study. As described, the normalized Witten index $\widetilde{W}$ is defined with the steady state distribution $\hat{\sigma}$, and thus the aim is to compute this distribution. 
As an example, we consider open system dynamics leading to a thermal distribution of the system in the steady state regime. The physical picture corresponds to the spin chain, governed by  $H_{\rm XXZ}$, defined in the main text, exchanging energy with the thermal reservoir at inverse temperature $\beta$.
This open system dynamics can be described using a variety of analytical and numerical techniques, including the Lindblad master equation approach, quantum collisional models, and Monte Carlo methods 
\cite{Strasberg2017, Barra2017}. 
The Monte Carlo Metropolis algorithm is an iterative method that is used to sample from a target distribution 
\cite{metropolis1953, myers2024}. 
This method consists of proposing updates to the states in each iteration, and then accepting or rejecting those updates based on the Metropolis filter. 

The first step in our method consists of finding the eigenstates $\{\psi\}$ and eigenvalues $\{E\}$ of the $H_{\rm XXZ}$. The next steps are outlined in Algorithm ~\eqref{algo}. For many-body thermalization, updates consist of transitions from the current eigenstate to another randomly chosen eigenstate. This dynamics ensures the ergodicity required for thermalization \cite{arisoy2019}. Additional details on the connection of this algorithm to collisional model approaches can be found in previous work \cite{myers2024}.
Note that the state in every iteration corresponding to one Monte Carlo run is a pure state, and the density distribution is constructed due to averaging over many such runs. The updating scheme of transitions between the eigenstates is common to both the methods that we use in our work - the grand-canonical averaging (GCA) method, and the quasi-grand canonical averaging (QGCA) method.  We discuss these methods and the equivalence between the two at low temperatures in the following subsections.

\begin{algorithm*}
\caption{Thermalization using Metropolis Algorithm}
\KwData{Initial eigenstate $\psi_i$, Number of thermalization iterations $num\_thermalization$, Number of Monte Carlo runs $num\_runs$}.
\For{$i \leftarrow 1$ \KwTo $num\_runs$} {
    Initialize system into some randomly chosen eigenstate\;
\For{$n \leftarrow 1$ \KwTo $num\_thermalization$}{
    Propose a jump to a new eigenstate $\psi'$\;
    Calculate the energy of state $\psi$ as 
    $E = \langle \psi | H | \psi \rangle$, and similar 
    for state $\psi'$. Compute the energy difference $\Delta E = E^{'} - E$.
    Calculate the acceptance ratio $\alpha = \min\left(1, {\rm exp}(-\beta \Delta E) \right)$\;
    Generate a random number $u$ from a uniform distribution $[0, 1]$\;
    \eIf{$u < \alpha$}{
        Accept the proposed eigenstate jump: $\psi \leftarrow \psi'$\;
    }{
        Reject the proposed eigenstate jump: $\psi$ remains unchanged\;
    }
    Record the current state $\psi$.
}
}
Calculate the thermal averages or occupation probabilities of the eigenstates by averaging over Monte Carlo runs.
\label{algo}
\end{algorithm*}

\subsection{Grand canonical approach}
As described in the main text, the SUSY spectrum of the XXZ model is comprised of eigenstates of different chain sizes. 
In the GCA, we consider the excitations coupling the eigenstate $\psi_i^{L}$ to $\psi_j^{L^{'}}$, where the superscript $L$ denotes the length, to obtain the thermal distribution over the eigenstates constituting the SUSY spectrum. As stated before, the state in $i$th iteration and the $p$th Monte Carlo run is a pure state, whose density matrix can be denoted as $\rho^{p}(t_i)$. In every iteration then, the updates to any randomly chosen eigenstate of a randomly chosen chain size are proposed, and the accept or reject criteria follows. Note that the eigenstate in every iteration need not always be present in the SUSY spectrum.  Therefore, we only compute $\widetilde{w}^{p}(t_i)$, which we refer as the contributory Witten index, when $\rho^{p}(t_i)$ is a state in the SUSY spectrum, and refer to the $p$th sample for the $i$th iteration as legitimate: $$\widetilde{w}^{p}(t_i) = {\rm Tr}\left[(-1)^{n_d}\rho^{p}(t_i)\right].$$ 
The normalized Witten index is then obtained by averaging over legitimate $N_{\rm run}$ Monte Carlo runs of the $i$th iteration resulting in
\begin{equation}
 \widetilde{W}(t_i) =\frac{1}{N_{\rm run}} \sum_{p = 1}^{N_{\rm run}} \widetilde{w}^{p}(t_i) 
\end{equation}
at $i$th iteration. 

\subsection{Quasi-grand canonical approach and proof of equivalence}
In the QGCA, we consider an ensemble of spin chains of all the lengths, exchanging energy with the ancilla drawn from the thermal reservoir corresponding to the transition frequencies of the chain. With these system-bath interactions, each chain individually equilibrates to the thermal state, and the updates proposing jumps across eigenstates of different lengths are not considered. 

We prove equivalence between the GCA and QGCA at low temperatures.
The Witten index in the steady state is given as 
\begin{equation}
 \widetilde{W}(t \rightarrow \infty) = \sum_{L=L_{\rm min}}^{L=L_{\rm max}}\sum_{i=1}^{N_L} \left(\frac{{\rm exp}(-\beta E_i)}{Z_L}\right)\left(-1\right)^{n_d(L)}.
 \label{witten_qgc}
\end{equation}
That is, the Witten index is computed by averaging over all the contributions from the length sectors in the domain $[L_{\rm min}, L_{\rm max}]$ and the prefactor is the thermal probability of the $i$th eigenstate of the $L$ sized chain. $N_L$ represents the number of eigenstates of the $L$ length chain that are in the SUSY spectrum for the given $N$, while $n_d(L) = N - L - 1$ is the number of down spins that these eigenstates have. The $L_{\rm min} (L_{\rm max})$ is the minimum(maximum) length in the $N$ sector. 
The above equation can be written as: 
\begin{eqnarray}
 \widetilde{W}(t \rightarrow \infty) = \sum_{L=L_{\rm min}}^{L=L_{\rm max}-1}&&(-1)^{k_L} \left(-1\right)^{n_0} \frac{1}{2}\left(\frac{1}{Z_L} - \frac{1}{Z_{L+1}}\right)\left[{\rm exp}(-\beta E_1) + {\rm exp}(-\beta E_2) + \ldots {\rm exp}(-\beta E_n)\right]  \nonumber \\
   && + m \left(-1\right)^{n_{\tilde{L}}}/Z_{\tilde{L}},
\end{eqnarray}
where $m = 0 (m = 1)$ when $N = 3j (N \ne 3j)$, and
$(E_1 < E_2 < \ldots E_n)$ are the ordered energies degenerate between the $L$ and the $L+1$ sector, $n_0 =  N - L_{\rm min}-1$ is the number of down spins for $L_{\rm min}$ , and $k_L = L_{\rm min} - L$. The nondegenerate $E=0$ mode, if present in the SUSY spectrum (when $m=1$), is assumed to be for the chain of length $\tilde{L}$. The difference in the inverse partition functions arises due to degenerate eigenstates with opposite parities. 
The XXZ model has the zero energy mode for every chain size \cite{yang_2004_nonlocal}, and thus in the low-temperature limit, $\beta \rightarrow \infty$, $Z_L \approx 1$. 
Therefore, $\frac{1}{Z_L} - \frac{1}{Z_{L+1}} = 0$ in this limit. This results in $\widetilde{W} = 0$ when $m=0$, and $W=(-1)^{\floor{N/3}}$ when $m=1$, thereby explaining the numerical results shown in Fig.~(3) in the main text.

\begin{figure}[h]
   \includegraphics[width=0.7\linewidth]{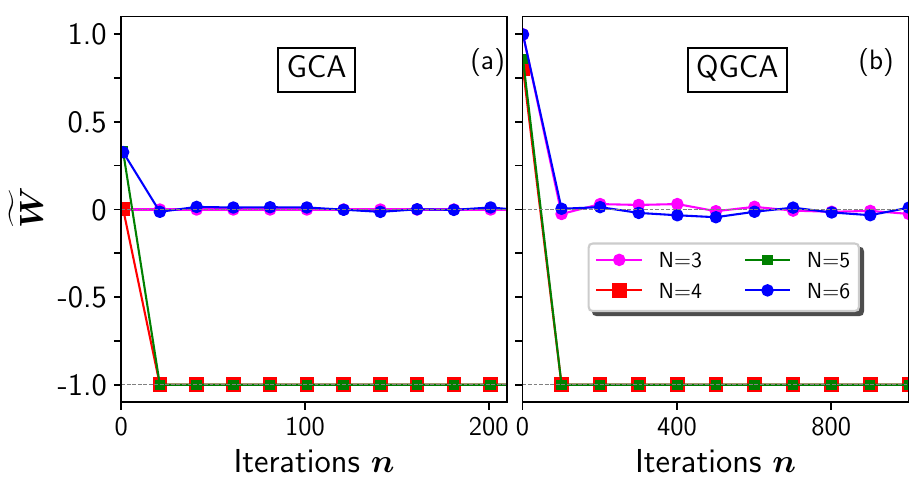}
   \caption{Panel a(b) illustrates $\widetilde{W}$ as a function of iterations $n$ of the system-bath interactions for different $N$ for $\beta=5$ computed with GCA (QGCA). The results are generated by averaging over $50000$ Monte Carlo runs. The largest error in the steady-state values is of the order $O(10^{-3})$ in (a) and $O(10^{-2})$ in (b). The $N=3$ (magenta) and $N=6$ (blue) cases average to zero, while $N=4$ (red) and $N=5$ (red) average to $-1$ in both GCA and QGCA.} 
    \label{wittn_iter_b5_nsmall} 
\end{figure}

We present the data for smaller values of $N$: $N = 3,4,5,6$, for $\beta=5$ in Fig.~\ref{wittn_iter_b5_nsmall}. The results with large $N$ values are shown in Fig. 3 in the main text. Since this temperature is less than the energy gap, the Witten index can be inferred from the QGCA.

\begin{figure}[h]
   \includegraphics[width=0.7\linewidth]{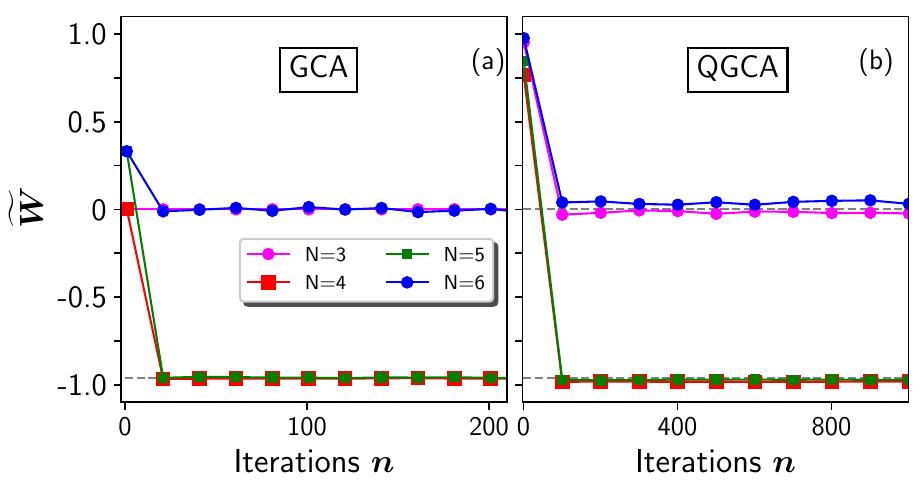}
   \caption{Panel a(b) illustrates $\widetilde{W}$ as a function of iterations $n$ of the system-bath interactions for different $N$ for $\beta=2$ computed with GCA (QGCA). The results are generated by averaging over $50000$ Monte Carlo runs. In (a), the $N=3$ (magenta) and $N=6$ (blue) cases average to zero. The GCA and QGCA yield similar results upto $N=5$ owing to the larger energy gap for smaller $N$ values. The largest error in the steady-state values is of the order $O(10^{-3})$ in (a) and $O(10^{-2})$ in (b). } 
    \label{wittn_iter_b2_nsmall} 
\end{figure}

\begin{figure}[h]
   \includegraphics[width=0.7\linewidth]{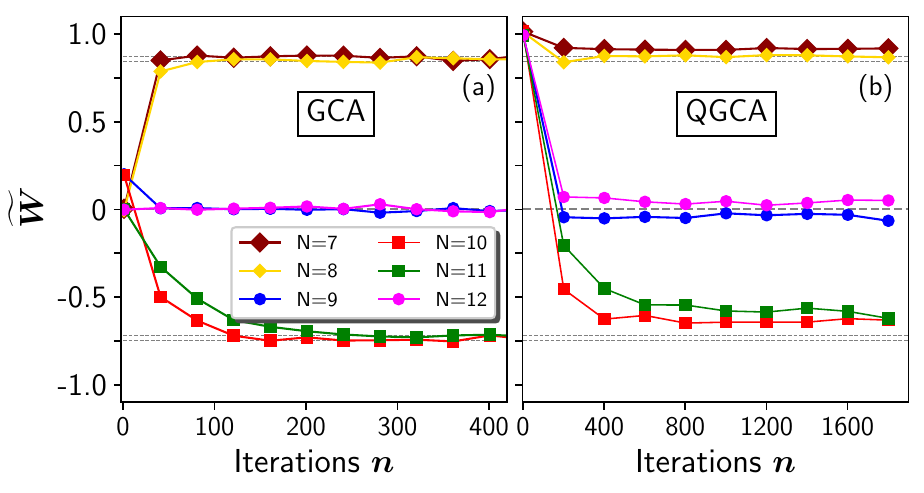}
   \caption{Panel a(b) illustrates $\widetilde{W}$ as a function of iterations $n$ of the system-bath interactions for different $N$ for $\beta=2$ computed with GCA (QGCA). In (a), the $N=9$ (blue) and $N=12$ (magenta) average to zero. Both approaches do not yield similar results, because the temperature is comparable to the gap. The results are generated by averaging over $50000$ Monte Carlo runs. The largest error in the steady-state values is of the order $O(10^{-3})$ in (a) and $O(10^{-2})$ in (b).} 
    \label{wittn_iter_b2} 
\end{figure}

However, when the temperature is higher than the energy gap, the equivalence breaks as $Z_{L} \neq Z_{L+1}$ and the degenerate eigenstates have biased weights, thus leading to deviations in $\widetilde{W}$, as shown in Fig.~\ref{wittn_iter_b2_nsmall} and Fig.~\ref{wittn_iter_b2}. The steady-state values obtained from the GCA are in excellent agreement with the thermal values, indicated by the dashed lines, in Fig.~\ref{wittn_iter_b2_nsmall}a and Fig.~\ref{wittn_iter_b2}a. However, $\widetilde{W}$ computed using the QGCA, shown in Fig.~\ref{wittn_iter_b2_nsmall}b and Fig.~\ref{wittn_iter_b2}b, does not agree with expected values. In particular, the results for larger $N$ (from $N=6$ onward) are not as expected, while the smaller $N$ results (upto $N = 5$) are in agreement with the expected values. This is because for larger $N$, the energy gap is comparatively smaller than the ones for smaller $N$ sectors. Thus for smaller $N$, the thermal fluctuations are suppressed by a large energy gap, while for larger $N$ sectors, they become prominent, leading to the breakdown in the approximation. Nevertheless, the equivalence at low temperatures implies that the study of the SUSY in the XXZ model can be done with QGCA which is the method of relevance for experiments. 

In order to understand the feasibility of QGCA with larger system sizes, we do finite-size scaling of the lowest energy gap by diagonalizing Eq.~(3) in the main text. We extract the lowest energy gap in the thermodynamic limit from the data from $L=4$ to $L=12$ sites. We observe that the gap decreases linearly and that the value in the thermodynamic limit is about $0.03 J$. This value sets a constraint on the energy scales required for the equivalence between the GCA and QGCA to be valid in the thermodynamic limit. However, for the XXZ model, the pattern in $\widetilde{W}$ can be extracted even with moderate system sizes, as it does not scale. Consequently, even small system simulations can reveal $\widetilde{W}$.

\subsection{Approximating GCA with QGCA}
In this subsection, we investigate the general conditions under which the GCA can be approximated with the QGCA. Consider a generic operator $\hat{O}$ for which we want to compute the grand canonical average. For the analysis, let us consider two system sizes, $L_1$ and $L_2$, and let us consider two eigenstates $\{\ket{\psi}_1,\ket{\psi}_2\} \Big(\{\ket{\psi}_3,\ket{\psi}_4\}\Big) $
with eigenvalues $\{\epsilon_1,\epsilon_2\} \Big(\{\epsilon_3,\epsilon_4\}\Big)$ for size $L_1 (L_2)$.  
Furthermore, let us denote $O_i = {}_{i}\bra{\psi}\hat{O} \ket{\psi}_i.$

The GCA of $\hat{O}$ would be 
\begin{eqnarray}
  &&\langle \hat{O} \rangle_{\rm GCA} = \frac{\sum_{i} O_{i} {\rm exp}(-\beta \epsilon_{i})}{Z_{\rm GCA}} \nonumber \\
  \label{O_GCA}
\end{eqnarray}

The grand canonical partition function would be  
$$Z_{\rm GCA} = {\rm exp}(-\beta \epsilon_{1}) + {\rm exp}(-\beta \epsilon_{2}) + {\rm exp}(-\beta \epsilon_{3}) + {\rm exp}(-\beta \epsilon_{4}).$$

The QGCA of $\hat{O}$ would be 
\begin{eqnarray}
    && \langle \hat{O} \rangle_{\rm QGCA} = \frac{1}{2}\Bigg(\frac{O_1 {\rm exp}(-\beta \epsilon_{1}) + O_2{\rm exp}(-\beta \epsilon_{2})}{Z_{L_1}} +
    \frac{O_3 {\rm exp}(-\beta \epsilon_{3}) + O_4{\rm exp}(-\beta \epsilon_{4})}{Z_{L_2}}\Bigg) \nonumber \\
    \label{O_QGCA}
\end{eqnarray}
where $Z_{L_1} (Z_{L_2})$ is the canonical partition function for size $L_1 (L_2)$.
Let us express $\epsilon_3 = \epsilon_1 + \lambda_1$ and 
$\epsilon_4 = \epsilon_2 + \lambda_2$, and then calculate 
the difference between $\langle \hat{O} \rangle_{\rm GCA}$ and $\langle \hat{O} \rangle_{\rm QGCA}$. 
Let us denote ${\rm exp}(-\beta \epsilon_{i}) = p_i$ and rewrite Eq.~\eqref{O_QGCA} as 
\begin{eqnarray}
   &&\langle \hat{O} \rangle_{\rm QGCA} =  \frac{1}{2}\Bigg(\frac{O_{1} p_{1} + O_{2} p_{2}}{p_{1} + p_{2}} + \frac{O_{3} p_{3} + O_{4} p_{4}}{p_{3} + p_{4}}\Bigg)  \nonumber \\
   &=&  \frac{1}{2}\Bigg(\frac{O_{1} p_{1} + O_{2} p_{2}}{p_{1} + p_{2}} + \frac{O_{3} p_{1} {\rm exp}(-\beta \lambda_1)+ O_{4} p_{2}{\rm exp}(-\beta \lambda_2)}{p_{1}{\rm exp}(-\beta \lambda_1) + p_{2}{\rm exp}(-\beta \lambda_2)}\Bigg). 
   \label{O_QGCA_1}
\end{eqnarray}
Let us denote, $$\frac{O_{3} p_{1} {\rm exp}(-\beta \lambda_1)+ O_{4} p_{2}{\rm exp}(-\beta \lambda_2)}{p_{1}{\rm exp}(-\beta \lambda_1) + p_{2}{\rm exp}(-\beta \lambda_2)} = f(\lambda_1,\lambda_2).$$
We now Taylor expand $f(\lambda_1,\lambda_2)$ about $(\lambda_1,\lambda_2) = (0,0)$ up to first order in $\lambda_1$ and $\lambda_2$, assuming they are small:
\begin{eqnarray}
   && f(\lambda_1,\lambda_2) \approx f(0,0) + \frac{\partial f}{\partial \lambda_1}\Big|_{\lambda_1 = 0} \lambda_1 + \frac{\partial f}{\partial \lambda_2}\Big|_{\lambda_2 = 0} \lambda_2 \nonumber \\
  &=& \frac{O_{3} p_{1} + O_{4} p_{2}}{p_{1} + p_{2}} + \beta p_1 p_2 (O_4-O_3)\Bigg[\frac{\lambda_1 {\rm exp}(-\beta \lambda_2)}{\big(p_1 + p_2 {\rm exp}(-\beta \lambda_2) \big)^2} - \frac{\lambda_2 {\rm exp}(-\beta \lambda_1)}{ \big( p_1 {\rm exp}(-\beta \lambda_1) + p_2 \big)^2}\Bigg].  
  \label{taylor}
\end{eqnarray}
Using Eq.~\ref{O_GCA}, Eq.~\ref{O_QGCA_1} and Eq.~\ref{taylor}, we can write 
\begin{eqnarray}
    \langle \hat{O} \rangle_{\rm QGCA} - \langle \hat{O} \rangle_{\rm GCA} &=& 
     \frac{\beta}{2} p_1 p_2 (O_4-O_3)\Bigg[\frac{\lambda_1 {\rm exp}(-\beta \lambda_2)}{\big(p_1 + p_2 {\rm exp}(-\beta \lambda_2) \big)^2} - \frac{\lambda_2 {\rm exp}(-\beta \lambda_1)}{ \big( p_1 {\rm exp}(-\beta \lambda_1) + p_2 \big)^2}\Bigg]   \nonumber \\
    &=&\frac{\beta}{2} {\rm exp}(-\beta (\epsilon_1 +\epsilon_2) ) (O_4 - O_3)\Bigg[\frac{\lambda_1 {\rm exp}(-\beta \lambda_2)}{\big(p_1 + p_2 {\rm exp}(-\beta \lambda_2) \big)^2} - \frac{\lambda_2 {\rm exp}(-\beta \lambda_1)}{ \big( p_1 {\rm exp}(-\beta \lambda_1) + p_2 \big)^2}\Bigg]
    \label{GCA_QGCA}
\end{eqnarray}
where we take $\langle \hat{O} \rangle_{\rm GCA}$ at the reference point $(\lambda_1,\lambda_2) = (0,0)$ to understand the deviation due to non-zero $\lambda_1$ and $\lambda_2$.
This suggests that if 
the spectrum is nearly the same for the two system sizes, ($\lambda_1 \approx 0$ and $\lambda_2 \approx 0$), $\langle \hat{O} \rangle_{\rm GCA} = \langle \hat{O}\rangle_{\rm QGCA}$. And, Eq.~\eqref{GCA_QGCA} quantifies the difference between these two averages.
The choice to consider only two energy levels in the spectrum for both the system sizes is justified by considering a temperature range where contributions from higher energy levels in the thermal averaging are negligible. Consequently, our analysis demonstrates that the equivalence between the GCA and QGCA holds for temperatures below which the spectra for the two system sizes are nearly identical. Although this analysis focuses on a representative case of two system sizes, it can be generalized to accommodate larger numbers of system sizes.
For the SUSY analysis of the XXZ model, the equivalence between the GCA and QGCA works particularly well. This is due to the presence of an $E=0$ mode for all system sizes.
In this case, the requirement for the spectra to be nearly identical across different sizes translates to a low-temperature condition. This conclusion is supported by our calculations, which show that the equivalence between the GCA and QGCA is valid for temperatures below the energy gap.

\section{Topological protection data}
We consider the absolute difference $|\widetilde{W}-\widetilde{W}_{\rm SUSY}|$ as a measure of deviation from the SUSY when the parameters in the Hamiltonian are tuned away from the SUSY point. We first discuss the case when $\Delta$ is varied across $\Delta=1$, and obtain an analytical expression for $|\widetilde{W}-\widetilde{W}_{\rm SUSY}|$ as a function of $\Delta$ by doing a Taylor expansion up to first order in $(\Delta-1)$ about $\Delta=1$. For the two cases, $N=3j$ and $N\ne 3j$, this gives 

\begin{equation}
    \begin{cases}
      |\widetilde{W}-\widetilde{W}_{\rm SUSY}|  = c_{N}\beta |\Delta-1| & \text{when $N = 3j$} \\
      |\widetilde{W}-\widetilde{W}_{\rm SUSY}|  = c_{N}\beta {\rm exp}(-\beta E_1) |\Delta - 1| & \text{when $N \ne 3j$}
    \end{cases}
    \label{susy_deviation}
\end{equation}
where $c_{N}$ is proportional to the slope of the degeneracy splitting as a function of $\Delta$, and $E_1$ is the lowest positive energy. The presence of the nondegenerate $E=0$ mode at the SUSY point, when $N\ne 3j$, leads to the exponential factor ${\rm exp}(-\beta E_1)$ that explains the topological protection when the thermal fluctuations are minimal compared to $E_1$, as shown in Fig.~4 in the main text.  Note that it is the rate of the degeneracy splitting $c_N$ that determines $|\widetilde{W}-\widetilde{W}_{\rm SUSY}|$. For $N=6$, the change in $\widetilde{W}$ is around $10 \%$ ($c_6 \approx 4\times10^{-2}$) for $\beta=5$. On the contrary, at same $\beta$, the variation in $\widetilde{W}$ for $N=9$ as a function of $\Delta$ is small (See Fig.~4 in the main text). This can be attributed to a relatively small $c_9 \approx 10^{-3}$, and is purely due to energetics as discussed in Sec.~\ref{perturbation_c9}, unrelated to the gap protection. To further validate that, as shown in Fig.~\ref{nonsusy_jvar_b5}, $\widetilde{W}$ changes relatively more for $N=9$ as a function of $J$, implying that $N=9$ is not topologically protected. Furthermore, the lowest energy gap decreases as system size increases, leading to larger deviations in larger $N$ sectors like $N=10$ and $N=11$ compared to lower sectors, as shown in Fig. 4 (II) inset in the main text.

\begin{figure}[h]
    \centering
    \includegraphics[width=0.7\linewidth]{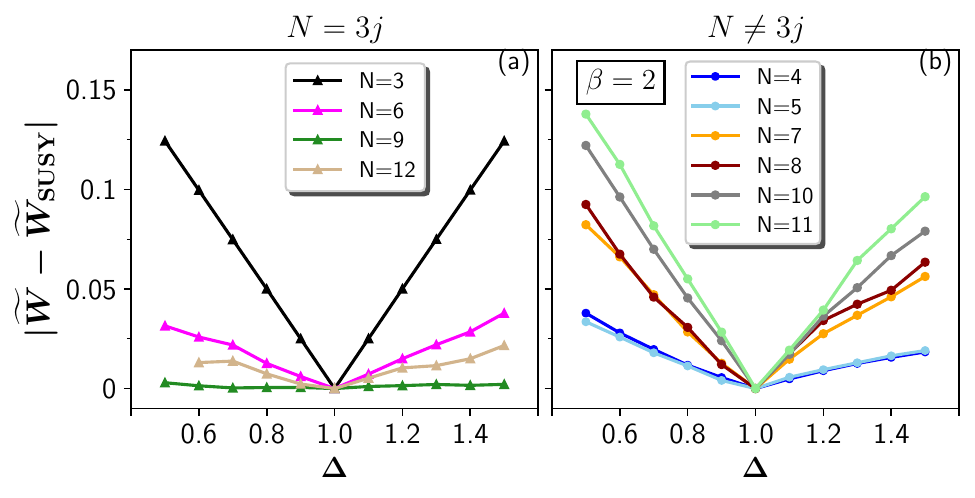}
    \caption{$|\widetilde{W} - \widetilde{W}_{\rm SUSY}|$ as a function of $\Delta$ traversed across the SUSY point, for $\beta = 2$, for two cases: $N=3j$ shown in (a) and $N \ne 3j$ shown in (b). These computations are done using GCA and the results are averaged over $50000$ samples. The characteristic error in both the cases is $O(10^{-2})$.  The thermal fluctuations are too large for panel (b) to show topological protection.}
    \label{nonsusy_b2}
\end{figure} 

However, this topological protection breaks down for temperatures near the energy gap, as shown in Fig.~\ref{nonsusy_b2}. The case demonstrated in Fig.~\ref{nonsusy_b2}a displays a linear deviation in $\widetilde{W}$. Furthermore, following Eq.~\eqref{susy_deviation}, the slope of this linear deviation is smaller as $\beta$ is smaller, compared to Fig.~4(I) in the main text. The relatively smaller variation in $\widetilde{W}$ for $N=9$ is due to a small $c_9$, as discussed in Sec.\ref{perturbation_c9} below. The $N\ne 3j$ case shown in Fig.~\ref{nonsusy_b2}b has a relatively larger deviation compared to Fig.~ 4(II) in the main text, implying that the robustness in $\widetilde{W}$ is lost.

\begin{figure}[h]
    \centering
    \includegraphics[width=0.7\linewidth]{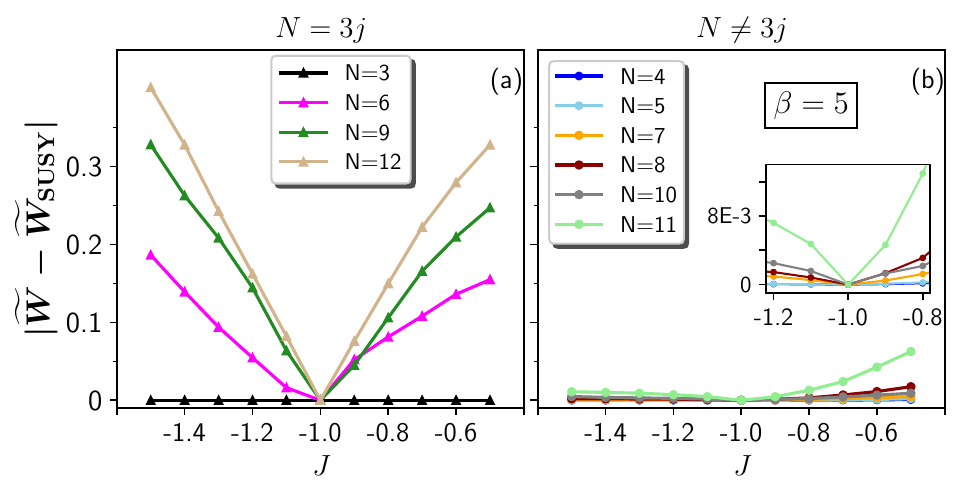}
    \caption{$|\widetilde{W} - \widetilde{W}_{\rm SUSY}|$ as a function of $J$ traversed across the SUSY point, for $\beta = 5$, for the two cases - $N=3j$ shown in (a) and $N \ne 3j$ shown in (b). The latter case featuring a nonzero $\widetilde{W}$ is protected due to the energy gap. The inset shows a zoomed picture of $|\widetilde{W}-\widetilde{W}_{\rm SUSY}|$ near the SUSY point. These computations are done using GCA and are averaged over $50000$ samples. The error in (a) is $O(10^{-2})$, while it is comparable to the point size in (b).} 
    \label{nonsusy_jvar_b5}
\end{figure}

We also investigate trend of $|\widetilde{W}-\widetilde{W}_{\rm SUSY}|$ when $J$ is varied across the SUSY point $J=-1$. We first discuss the results obtained by GCA at low temperatures. Fig.~\ref{nonsusy_jvar_b5} illustrates $|\widetilde{W}-\widetilde{W}_{\rm SUSY}|$, computed using GCA, as a function of $J$. In Fig.~\ref{nonsusy_jvar_b5}a, the results for $N=3j$ case are shown. Except for $N=3$, $|\widetilde{W}-\widetilde{W}_{\rm SUSY}|$ has a linear trend, inline with the prediction of Eq.~\eqref{susy_deviation}. For $N=3$, the two eigenstates constituting the SUSY spectrum are $\ket{\downarrow}$ and $\ket{\uparrow\uparrow}$, and their energies do not depend on $J$, thereby $\widetilde{W}-\widetilde{W}_{\rm SUSY} = 0$ for all $J$. 
For the case where $N\ne 3j$, the difference $|\widetilde{W}-\widetilde{W}_{\rm SUSY}|$ is significantly smaller, as illustrated in Fig.~\ref{nonsusy_jvar_b5}b, indicating that $\widetilde{W}$ is topologically protected. 
Furthermore, the inset shows that the deviation in $\widetilde{W}$ for the higher $N$ sectors is relatively larger than the lower $N$ sectors, due to the smaller energy gap for higher $N$ sectors.

\begin{figure}[h]
    \centering
    \includegraphics[width=0.7\linewidth]{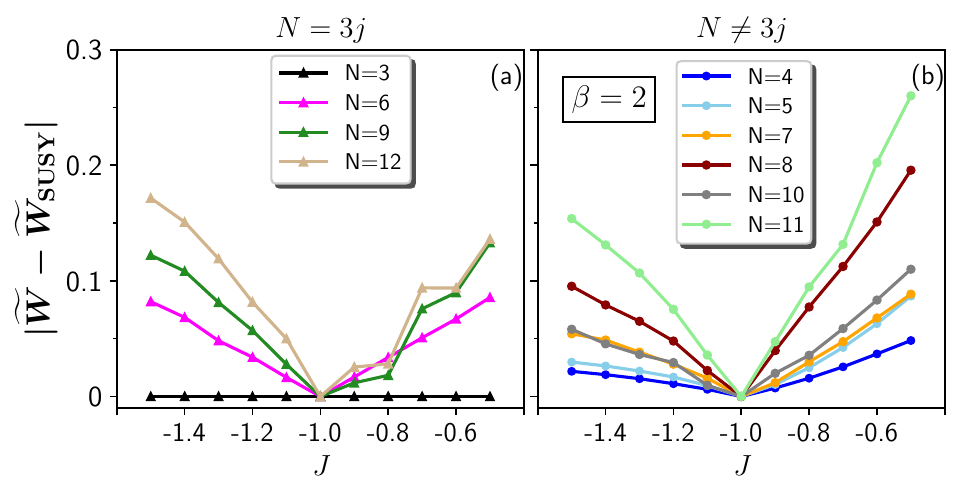}
    \caption{$|\widetilde{W} - \widetilde{W}_{\rm SUSY}|$ as a function of $J$ traversed across the SUSY point, for $\beta = 2$, for two cases - $N=3j$ shown in (a) and $N \ne 3j$ shown in (b). There is a significant deviation in $\widetilde{W}$, away from the SUSY, due to high thermal fluctuations.
    These computations are done using GCA and are averaged over $50000$ samples. The error bars are $O(10^{-2}$) in both cases.} 
    \label{nonsusy_jvar_b2}
\end{figure} 

\begin{figure}[h]
    \centering
    \includegraphics[width=0.5\linewidth]{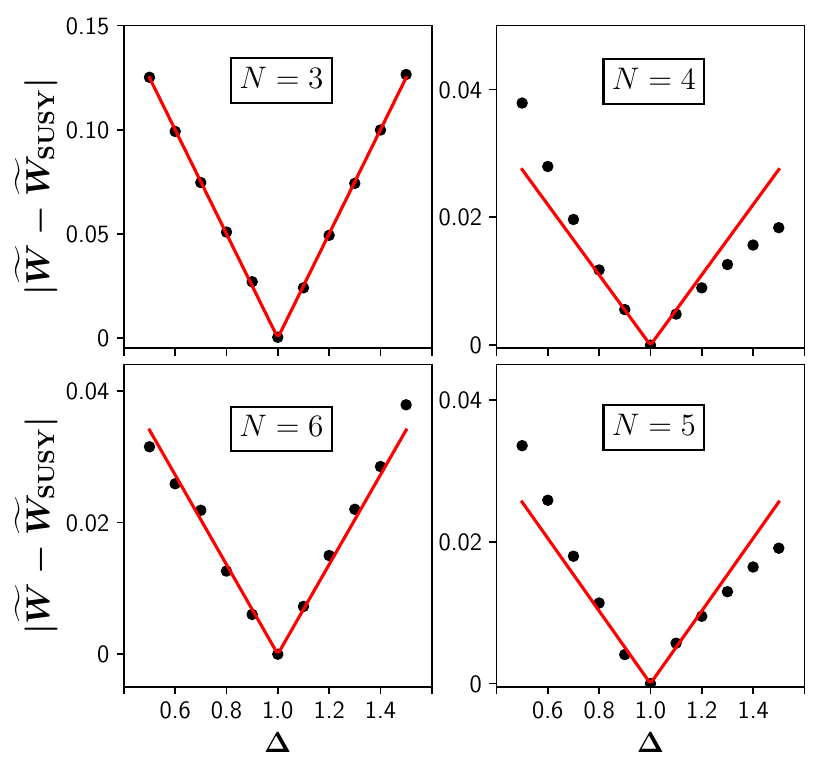}
    \caption{The comparison between the numerical data for $|\widetilde{W} - \widetilde{W}_{\rm SUSY}|$, as a function of $\Delta$, with the analytical expressions given in Eq.~\eqref{susy_deviation}, at $\beta=2$, for $N=3$ to $N=6$. The plots reveal an excellent agreement of the numerical data points, shown in black, with the analytical curves in red.} 
    \label{b2_fits}
\end{figure}

The high-temperature behavior of $|\widetilde{W}-\widetilde{W}_{\rm SUSY}|$ as a function of $J$ is shown in Fig.~\ref{nonsusy_jvar_b2}.
Compared to Fig.~\ref{nonsusy_jvar_b5}b, $\widetilde{W}$ displays 
enhanced behavior in Fig.~\ref{nonsusy_jvar_b2}b, indicating that the robustness is lost. The plots of $|\widetilde{W}-\widetilde{W}_{\rm SUSY}|$ as a function of $\Delta$ and $J$ across the SUSY point thus reveal topological protection at low temperatures.  We also compare our numerical results against the analytical expressions in Eq.~\eqref{susy_deviation}, and see excellent agreement between the two, as demonstrated in Fig.~\ref{b2_fits} for representative cases of $N=3$ to $N=6$.

We now illustrate the  behavior of 
$|\widetilde{W} - \widetilde{W}_{\rm SUSY}|$ computed from the QGCA at $\beta=5$. At this temperature, QGCA is equivalent to GCA at the SUSY point.
\begin{figure}[h]
    \centering
    \includegraphics[width=0.7\linewidth]{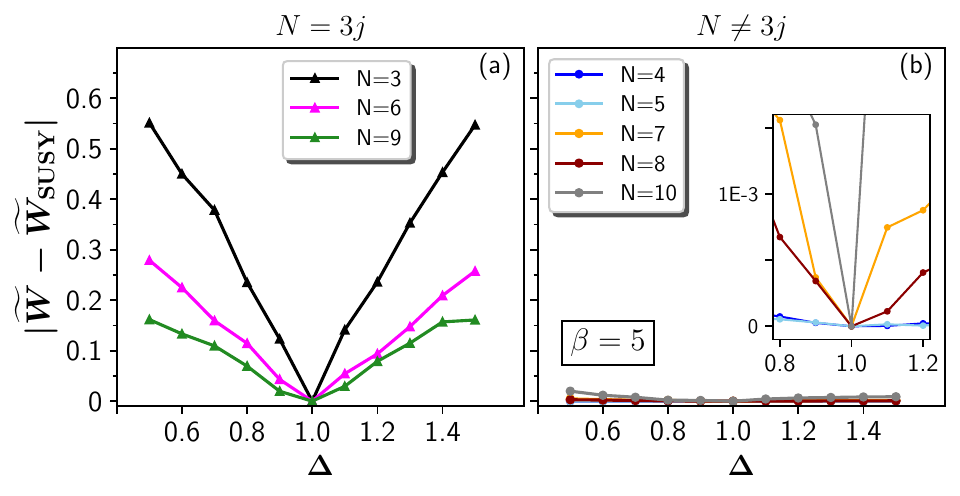}
    \caption{$|\widetilde{W} - \widetilde{W}_{\rm SUSY}|$ as a function of $\Delta$ traversed across the SUSY point, for $\beta = 5$, computed using QGCA, for two cases - $N=3j$ shown in (a), and $N \ne 3j$ shown in (b). The results obtained using this approach are qualitatively similar to the ones obtained using GCA. We notice the robustness of $\widetilde{W}$ against the SUSY breaking perturbations at this temperature. The inset in (b) reveals that deviations in $\widetilde{W}$ are suppressed by orders by magnitude compared to the deviation observed in subplot (a). All the results are generated by averaging over $50000$ Monte Carlo samples.}
    \label{nonsusy_qcan}
\end{figure}
We first plot $|\widetilde{W} - \widetilde{W}_{\rm SUSY}|$ as a function of 
$\Delta$, in Fig.~\ref{nonsusy_qcan}. For the $N=3j$ case, we observe significant deviations (with a nearly linear dependence) in $\widetilde{W}$ from the $\widetilde{W}_{\rm SUSY}$. For the $N \ne 3j$ case, we notice topological protection of $\widetilde{W}$ and $|\widetilde{W} - \widetilde{W}_{\rm SUSY}|$ is negligible ($\approx O(10^{-2}$)). These results are qualitatively similar to the results computed with GCA, further validating the equivalence between the two approaches at low temperatures. We have shown the calculations only up to the $N=10$ sector to avoid larger computations. A similar behavior of $|\widetilde{W} - \widetilde{W}_{\rm SUSY}|$ as a function of $J$ can be seen in Fig.~\ref{nonsusy_qcan_jvar}.
In both Fig.~\ref{nonsusy_qcan} and Fig.~\ref{nonsusy_qcan_jvar}, the error bars are $O(10^{-2})$ for $N=3j$ and $O(10^{-4})$ for $N \ne 3j$.
All of these results suggest that the 
$E=0$ mode (when $N \ne 3j$) leads to the topological protection of 
$\widetilde{W}$ at low temperatures.

\begin{figure}[h]
    \centering
    \includegraphics[width=0.7\linewidth]{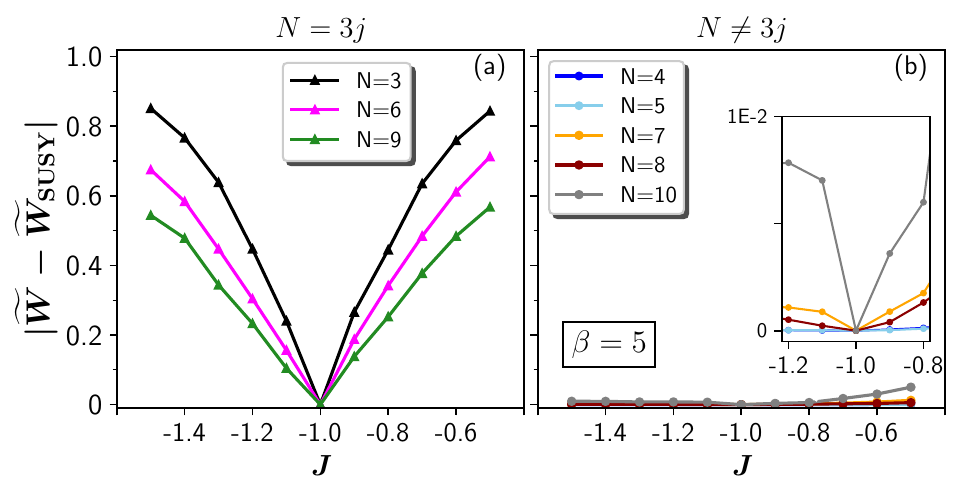}
    \caption{$|\widetilde{W} - \widetilde{W}_{\rm SUSY}|$ as a function of $J$ traversed across the SUSY point, for $\beta = 5$, computed using QGCA, for two cases - $N=3j$ shown in (a), and $N \ne 3j$ shown in (b). The results are generated by averaging over $50000$ Monte Carlo samples. The $\widetilde{W}$ is robust against the SUSY breaking perturbations at this temperature. The inset in (b) illustrates the behavior close to the $J=-1$.}
    \label{nonsusy_qcan_jvar}
\end{figure}

\section{Perturbative explanation of $c_9$}
\label{perturbation_c9}
As discussed previously, we have argued that the small value of $c_9$ leads to a vanishingly small variation in $\widetilde{W}$ for $N=9$ when $\Delta$ is varied across the SUSY point (See Fig.~4 in the main text and Fig.~\ref{nonsusy_b2}). This small value of $c_9$ is attributed to the nature of the eigenstates in the SUSY spectrum for $N=9$. As discussed previously, $N=9$ has two non-zero degenerate eigenstates of lowest energy, at the SUSY point. In this section, we shall estimate the change in the energies of these two eigenstates, when $\Delta$ is varied across the SUSY point, employing the first-order time-independent perturbation theory, for the two cases: $N=9$ and $N=12$ for comparison. Expanding the parameter $\Delta$ about the SUSY point while keeping other parameters at their SUSY value, i.e, $(J,\Delta, h) =  (-1,1 + d\Delta, 1/2)$, we decompose the XXZ Hamiltonian into the unperturbed Hamiltonian $\hat{H}_0$ and the perturbation $\hat{H}_1$ as 
\begin{eqnarray}
      \hat{H}_{0}  &=& \sum_{i=1}^{L-1} \left[-(\hat{S}_i^+ \hat{S}_{i+1}^- + {\rm H.c. } ) + \hat{S}_i^{z} \hat{S}_{i+1}^z\right] -  \frac{1}{2}\left(\hat{S}_1^z + \hat{S}_L^z \right) + \left(3L - 1 \right)/4, \nonumber \\
      \hat{H}_1 &=& d\Delta\sum_{i=1}^{L-1} \hat{S}_i^{z} \hat{S}_{i+1}^z .   
\end{eqnarray}
In other words, the variation in $\Delta$ about the SUSY point is considered as a perturbation. 
We then estimate the energy correction of the two lowest energy eigenstates
as a function of $d\Delta$, and then take their difference. This difference is the degeneracy splitting $\delta_N$, 
and the rate of change of $\delta_N$ when $d\Delta$ is varied yields $c_N$. We compare the estimate of $\delta_N$, obtained from the perturbative analysis, with the exact-diagonalization (ED) calculation away from the SUSY point. Note that we have utilized the ED for computing the eigenvectors at the SUSY point that are then required in perturbative estimates. 
\begin{figure}[H]
    \centering
    \includegraphics[width=0.5\linewidth]{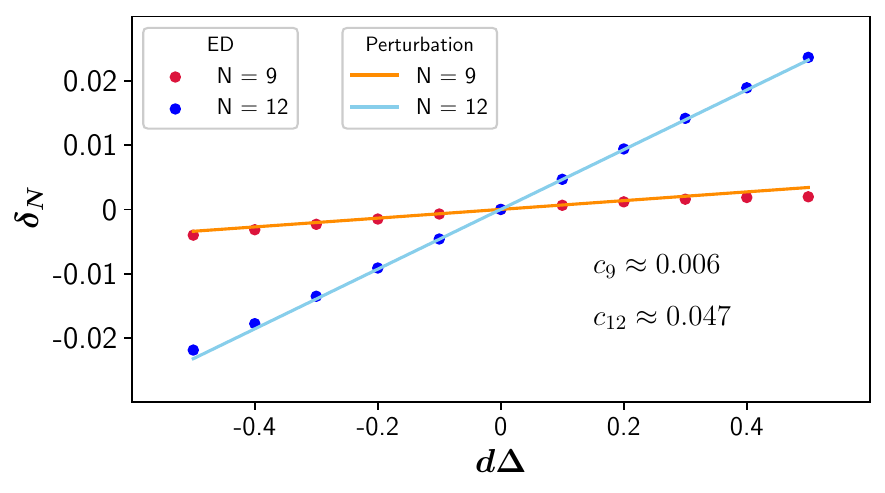}
    \caption{Perturbative estimate of the degeneracy splitting $\delta_N$ between the lowest energy eigenstates in the SUSY spectrum for $N=9$ and $N=12$, when $\Delta$ is varied away from the SUSY point. 
    The red (blue) filled circles indicate the data from the exact-diagonalization calculation for $N=9 (N=12)$, while the brown (cyan) solid lines indicate the data computed using the first-order perturbation theory. The plot indicates an excellent agreement between the ED data and the perturbation theory estimates. 
    The slope of the plot is a measure of $c_N$, and we notice that $c_{12}$ is an order of magnitude larger than the $c_9$, thereby leading to a vanishingly small $|\widetilde{W} - \widetilde{W}_{\rm SUSY}|$ for the later case. }
    \label{perturbation_fig}
\end{figure}

In Fig.~\ref{perturbation_fig}, $\delta_N$ is plotted as a function of $d\Delta$ using the ED data and the perturbative estimates. The figure reveals that the first-order perturbation theory gives excellent predictions of $\delta_N$, as these estimates agree perfectly well with the ED results. More importantly, it highlights that the slope for $N=9$ is smaller in comparison to the slope for $N=12$, which implies that the two lowest energy eigenstates for $N=9$ are less sensitive to the variations in $\Delta$. This explains why $|\widetilde{W} - \widetilde{W}_{\rm SUSY}|$ is small for $N=9$ in comparison to other $N=3j$ cases. This perturbative technique can be extended to other $N$ values and when $J$ is varied across the SUSY point to estimate the $c_N$, and understand the nature of the degeneracy splitting. 

\bibliography{witten_ind_sup}